# Forecasting Uncertain Events with Small Groups[*]

Kay-Yut Chen, Leslie R. Fine and Bernardo A. Huberman

HP Laboratories

Palo Alto, CA 94304

June 25, 2001

## Abstract

We present a novel methodology for predicting future outcomes that uses small numbers of individuals participating in an imperfect information market. By determining their risk attitudes and performing a nonlinear aggregation of their predictions, we are able to assess the probability of the future outcome of an uncertain event and compare it to both the objective probability of its occurrence and the performance of the market as a whole. Experiments show that this nonlinear aggregation mechanism vastly outperforms both the imperfect market and the best of the participants.

---

## Introduction

The prediction of the future outcomes of uncertain situations is both an important problem and a guiding force behind the search for the regularities that underlie natural and social phenomena. While in the physical and biological sciences the discovery of strong laws has enabled the prediction of future scenarios with uncanny accuracy, in the social sphere no such accurate laws are known. To complicate matters further, in social groups the information relevant to predictions is often dispersed across people, making it hard to identify and aggregate it.

In the social arena, economists have long articulated the belief that markets efficiently collect and disseminate information [1]. In particular, rational expectations theory tells us that markets have the capacity not only to aggregate information held by individuals, but also to convey it via the price and volume of assets associated with that information. Therefore, a possible methodology for the prediction of future outcomes is the construction of markets where the asset is information rather than a physical good. Laboratory experiments have determined that these markets do indeed have the capacity to aggregate information in this type of setting [2, 3, 4, 5].

Information markets generally involve the trading of state-contingent securities. If these markets are large enough and properly designed, they can be more accurate than other techniques for extracting diffuse information, such as surveys and opinions polls. There are problems however, with information markets, as they tend to suffer from information traps [6, 7], illiquidity [8], manipulation [9, 10], and lack of equilibrium [11, 12][1]. These problems are exacerbated when the groups involved are small and not very experienced at playing in these markets. Even when possible, proper market design is very expensive, fragile, and context-specific.

---

[1] Notable exceptions: The Iowa Electronic Market [13] has shown that political events can be accurately predicted using markets when they are large enough. Their predictions have consistently been more accurate than those resulting from major news polls. Additionally, recent work by Pennock, Lawrence, Giles and Nielsen [14] show that the Hollywood Stock Exchange (HSX) does a remarkable job of predicting box office revenues and Oscar winners. However, both of these institutions have many traders, while we focus on systems with small number of participants (fewer than 15).

In spite of these obstacles, it is worth noting that certain participants in information markets can have either superior knowledge of the information being sought, or are better processors of the knowledge harnessed by the information market itself. By keeping track of the profits and final holdings of the members, one can determine which participants have these talents, along with their risk attitudes.

In this paper we propose a method of harnessing the distributed knowledge of a group of individuals by using a two-stage mechanism. In the first stage, an information market is run among members of the group in order to extract risk attitudes from the participants, as well as their ability at predicting a given outcome. This information is used to construct a nonlinear aggregation function that allows for collective predictions of uncertain events. In the second stage, individuals are simply asked to provide forecasts about an uncertain event, and they are rewarded according the accuracy of their forecasts. These individual forecasts are aggregated using the nonlinear function and used to predict the outcome. As we show empirically, this nonlinear aggregation mechanism vastly outperforms both the imperfect market and the best of the participants.

**Information Aggregation Mechanism Design**

In order to construct the aggregation function, we first notice that in ideal settings, it is easy to compute the true posterior probabilities using Bayes' rule. If individuals receive independent information conditioned on the true outcome, their prior beliefs are uniform (no other information is available other than the event sequence), and they each report the true posterior probabilities given their information, then the probability of an outcome *s*, conditioned on all of their observed information *I*, is given by:

$$P(s \mid I) = \frac{p_{s_1} p_{s_2} \cdots p_{s_N}}{\sum_{\forall s} p_{s_1} p_{s_2} \cdots p_{s_N}} \qquad (1)$$

where $p_{si}$ is the probability that individual *i* (*i=1…N*) assigns to outcome *s* (please see Appendix 1 for a discussion). This result allows us simply to take the individual predictions, multiply them together, and normalize them in order to get an aggregate probability distribution. The issue becomes how to design a mechanism that elicits truthful reporting from individuals. We demonstrate in Appendix 2 that the following mechanism will induce risk neutral utility maximizing individuals to report their prior probabilities truthfully. We ask each player to report a vector of perceived state-probabilities, $\{q_1, q_2, ... q_N\}$ with the constraint that the vector sums to one. Then the true state $x$ is revealed and each player paid $c_1 + c_2 * log(q_x)$, where $c_1$ and $c_2$ are positive numbers.

While this very simple method might seem to aggregate dispersed information well, it suffers from the fact that, due to their risk attitude, most individuals do not necessarily report their true posterior probabilities conditioned on their information. In most realistic situations, a risk averse person will report a probability distribution that is flatter than her true beliefs as she tends to spread her bets among all possible outcomes. In the extreme case of risk aversion, an individual will report a flat probability distribution regardless of her information. In this case, no predictive information is revealed by her report. Conversely, a risk-loving individual will tend to report a probability distribution that is more sharply peaked around a particular prediction, and in the extreme case of risk loving behavior a subject's optimal response will be to put all his weight on the most probable state according to his observations. In this case, his report will contain some, but not all the information contained in his observations.

In order to account for both the diverse levels of risk aversion and information strengths, we add a stage to the mechanism. Before individuals are asked to report their beliefs, they participate in an information market designed to elicit their risk attitudes and other relevant behavioral information. This information market is driven by the same information structure in the reporting game. We use information markets to capture the behavioral information that is needed to derive the correct aggregation function. Note that, although the participant pool is too small for the market to act perfectly efficiently, it is a powerful enough mechanism to help us illicit the needed information.

The nonlinear aggregation function that we constructed is of the form:

$$P(s \mid I) = \frac{p_{s_1}^{\beta_1} p_{s_2}^{\beta_2} ... p_{s_N}^{\beta_N}}{\sum_{\forall s} p_{s_1}^{\beta_1} p_{s_2}^{\beta_2} ... p_{s_N}^{\beta_N}} \qquad (2)$$

where $\beta_i$ is the exponent assigned to individual $i$. The role of $\beta_i$ is to help recover the true posterior probabilities from individual $i$'s report. The value of $\beta$ for a risk neutral individual is one, as he should report the true probabilities coming out of his information. For a risk averse individual, $\beta_i$ is greater than one so as to compensate for the flat distribution that he reports. The reverse, namely $\beta_i$ smaller than one, applies to risk loving individuals. In terms of both the market performance and the individual holdings and risk behavior, a simple functional form for $\beta_i$ is given by

$$\beta_i = r(V_i/\sigma_i)c \qquad (3)$$

where $r$ is a parameter that captures the risk attitude of the whole market and is reflected in the market prices of the assets, $V_i$ is the utility of individual $i$, and $\sigma_i$ is the variance of his holdings over time. $c$ is a normalization factor so that if $r=1$, $\sum_i \beta_i$ equals number of individuals. Thus the problem lies in the actual determination of both the risk attitudes of the market as a whole and of the individual players.

To do so, notice that if the market is perfectly efficient then the sum of the prices of the securities should be exactly equal to the payoff of the winning security. However, in the thin markets characterized here, this efficiency condition was rarely met. Moreover, although prices that do not sum to the winning payoff indicate an arbitrage opportunity, it was rarely possible to realize this opportunity with a portfolio purchase (once again, due to the thinness of the market). However, we can use these facts to our advantage. If the sum of the prices is below the winning payoff, then we can infer that the market is risk-averse, while if the price is above this payoff then the market exhibits risk-loving behavior. Thus, the ratio of the winning payoff to the sum of the prices provides a proxy for the risk attitude of the market as a whole.

The ratio of value to risk, $(V_i/\sigma_i)$, captures individual risk attitudes and predictive power. An individual's value $V_i$ is given by the market prices multiplied by his holdings, summed over all the securities. As in portfolio theory [15], his amount of risk can be measured by the variance of his values using normalized market prices as probabilities of the possible outcomes.

**Experimental Design**

In order to test this mechanism we conducted a number of experiments at Hewlett-Packard Laboratories, in Palo Alto, California. The subjects were undergraduate and graduate students at Stanford University and knew the experimental parameters discussed below, as they were part of the instructions and training for the sessions. The five sessions were run with eight to thirteen subjects in each.

The two-stage mechanism was implemented in a laboratory setting. Possible outcomes were referred to as "states" in the experiments. There were 10 possible states, A through J, in all the experiments. Each had an Arrow-Debreu[2] state security associated with it. The information available to the subjects consisted of observed sets of random draws from an urn with replacement. After privately drawing the state for the ensuing period, we filled the urn with one ball for each state, plus an additional two balls for the just-drawn true state security. Thus it is slightly more likely to observe a ball for the true state than others.

The amount of information given to the subjects was controlled by letting them observe different number of draws from the urn. Three types of information structures were used to ensure that the results obtained were robust. In the first treatment, each subject received three draws from the urn, with replacement. In the second treatment, half of the subjects received five draws with replacement, and the other half received one. In a third treatment, half of the subjects received a random number of draws (averaging three, and also set such that the total number of draws in the community was 3N) and the other half received three, again with replacement.

The information market we constructed consisted of an artificial call market in which the securities were traded. The states were equally likely and randomly drawn. If a state occurred, the associated state security paid off at a value of 1,000 francs[3]. Hence, the expected value of any given security, a priori, was 100 francs. Subjects were provided with some securities and francs at the beginning of each period.

---

[2] These securities have lottery-like properties, and they pay off one unit contingent on the positive outcome of an event linked to that security, and zero otherwise.

Each period consisted of six rounds, lasting 90 seconds each. At the end of each round, the bids and asks were gathered and a market price and volume was determined. The transactions were then completed and another call round began. At the end of six trading rounds the period was over, the true state security was revealed, and subjects were paid according to the holdings of that security. This procedure was then repeated in the next period, with no correlation between the states drawn in each period.

In the second-stage, every subject played under the same information *structure* as in the first stage, although the draws and the true states were independent from those in the first. Each period they received their draws from the urn and 100 tickets. They were asked to distribute these tickets across the 10 states with the constraint that all 100 tickets must be spent each period and that at least one ticket is spent on each state. Since the fraction of tickets spent determines $p_{si}$, this implies that $p_{si}$ is never zero. The subjects were given a chart that told them how many francs they would earn upon the realization of the true state as a function of the number of tickets spent on the true state security. The payoff is a linear function of the log of the percentage of tickets placed in the winning state(Please see Appendix 2 for a discussion of the payoff function). The chart the subjects received showed the payoff for every possible ticket expenditure, and an excerpt from the chart is shown below.

Table 1: Payoff Chart for Reporting Game

| Number of Tickets | Possible Payoff | Number of Tickets | Possible Payoff |
|---|---|---|---|
| 1 | 33 | 50 | 854 |
| 10 | 516 | 60 | 893 |
| 20 | 662 | 70 | 925 |
| 30 | 747 | 80 | 953 |
| 40 | 808 | 90 | 978 |

---

[3] An experimental currency, exchanged for dollars at the end of the experiment according to an announced exchange rate.

**Procedural Overview**

A total of five experiments were conducted. The number of subjects in the experiments ranged from eight to thirteen. The speed of the experiments depended on how fast the subjects were making their decisions, the length of the training sessions and a number of other variables. Therefore, we have completed different number of periods in different experiments. The following table provides a summary.

Table 2: Summary of Experiments

| Experiment Number | Number of Subjects | Number of Call Market Periods | Number of Rounds of Reporting Game | Information Structure |
|---|---|---|---|---|
| 1 | 13 | 3 | 7 | 3 draws |
| 2 | 9 | 6 | 18 | 3 draws |
| 3 | 11 | 7 | 29 | 5 draws for 6 subjects / 1 draw for 5 subjects |
| 4 | 8 | 7 | 25 | 5 draws for 4 subjects / 1 draw for 4 subjects |
| 5 | 10 | 10 | 30 | Random for 5 subjects / 3 draws for 5 subjects |

**Analysis**

Notice that if the aggregation mechanism were perfect, the probability distribution of the states would be as if one person had seen all of the information available to the community. Therefore, the probability distribution conditioned on all the information acts as a benchmark to which we can compare alternative aggregation mechanisms. In order to compute it, recall that there are twelve balls in the information urn, three for the true state and one for each of the other nine states. Using Bayes' rule one obtains the omniscient probability distribution, i.e.

$$P(s|O) = \frac{\left(\frac{3}{12}\right)^{\#(s)}\left(\frac{1}{12}\right)^{\#(\bar{s})}}{\sum_{\forall s}\left(\frac{3}{12}\right)^{\#(s)}\left(\frac{1}{12}\right)^{\#(\bar{s})}} \quad (4)$$

where *s* denotes the states, *O* is a string of observations, *#(s)* is the number of draws of the state *s* in the string, and $\#(\bar{s})$ is the number of draws of all other states.

Once this benchmark is created, the next step is to find a measure to compare probabilities provided by different aggregation mechanisms to this benchmark. The obvious measure to use is the Kullback-Leibler measure, also known as the relative entropy. The Kullback-Leibler measure of two probability distributions p and q is given by:

$$KL(p,q) = E_p\left(\log\left(\frac{p}{q}\right)\right) \quad (5)$$

where p is the "true" distribution. In the case of finite number of discrete states, the above equation (4) can be rewritten as:

$$KL(p,q) = \sum_s p_s \log\left(\frac{p_s}{q_s}\right) \quad (5)$$

It can be shown that a) KL(p,q)=0 if and only if the distribution p and q are identical, and b) KL(p,q)≥0. A smaller Kullback-Leibler number indicates that two probabilities are closer to each other.

Furthermore, the Kullback-Leibler measure of the joint distribution of multiple independent events is the sum of the Kullback-Leibler measures of the individual events. Since periods within an experiment were independent events, the sum or average (across periods) of Kullback-Leibler measures is a good summary statistics of the whole experiment.

**Results**

Three information aggregation mechanisms were compared to the benchmark distribution given by Eq. (4) by the use of the Kullback-Leibler measure. In addition, we also report the K-L measures of the "no information" prediction (uniform distribution over all the possible states) and the predictions of the best individual.

The "no information" prediction serves as the first baseline to determine if any information is contained in the predictions of the mechanisms. If a mechanism is really aggregating information, then it should be doing at least as well as the best individual. Predictions of the best individual serve as the second baseline, which helps us to determine if information aggregation indeed occurred in the experiments.

The first of the three information aggregation mechanisms is the market prediction. The market prediction was calculated using the last traded prices of the assets. We used the last traded prices rather than the current round's price because sometimes there was no trade in a given asset in a given round. From these prices, we inferred a probability distribution on the states.

The second and the third mechanisms are the simple aggregation function given by the risk neutral formula of Eq. (1), and the market-based nonlinear aggregation function of Eq. (2).

The results are shown in Table 3. The entries are the average values and standard deviations (in parentheses) of the Kullback-Leibler number [16], which was used to characterize the difference between the probability distributions coming out of a given mechanism and the omniscient probability.

Table 3: Kullback-Leibler Numbers, by Experiment

| No Information | Market Prediction | Best Player | Simple Aggregation Function | Nonlinear Aggregation Function |
|---|---|---|---|---|
| 1.977 (0.312) | 1.222 (0.650) | 0.844 (0.599) | 1.105 (2.331) | 0.553 (1.057) |
| 1.501 (0.618) | 1.112 (0.594) | 1.128 (0.389) | 0.207 (0.215) | 0.214 (0.195) |
| 1.689 (0.576) | 1.053 (1.083) | 0.876 (0.646) | 0.489 (0.754) | 0.414 (0.404) |
| 1.635 (0.570) | 1.136 (0.193) | 1.074 (0.462) | 0.253 (0.325) | 0.413 (0.260) |
| 1.640 (0.598) | 1.371 (0.661) | 1.164 (0.944) | 0.478 (0.568) | 0.395 (0.407) |

As can easily be seen, the nonlinear aggregation function worked extremely well in all the experiments. It resulted in significantly lower Kullback-Leibler numbers than the no information case, the market prediction, and the best a single player could do. In fact, it performed almost three times as well as the information market. Furthermore, the nonlinear aggregation function exhibited a smaller standard deviation than the market prediction, which indicates that the quality of its predictions, as measured by the Kullback-Leibler number, is more consistent than that of the market. In three of five cases, it also offered substantial improvements over the simple aggregation function.

The results displayed in the second column show that the market was not sufficiently liquid to aggregate information properly, and it was only marginally better than the a priori no information case.  In almost all cases, the best player in the reporting game conveyed more information about the probability distribution than the market did. However, even in situations where the market performs quite poorly, it does provide some information, enough to help us construct an aggregation function with appropriate exponents.

All these results are illustrated in Figure 1, we show the probability distributions generated by the market mechanisms, the best individual in a typical experiment, the nonlinear aggregation function, as well as the omniscient probability distribution generated by Equation (4) [4].  Notice that the nonlinear aggregation function exhibits a functional form very similar to the omniscient probability, and with low variance compared to the other mechanisms. This is to be contrasted with the market prediction, which exhibits information traps at state I and F, and a much larger variance.

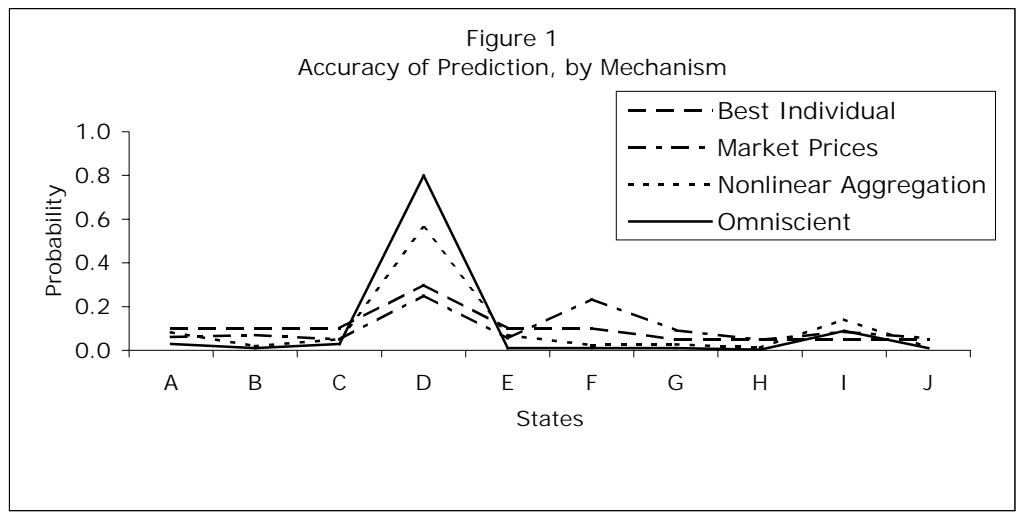

---

[4] While different, independent events are used for the market stage and the reporting stage, we found one period in both stages that contained the exact same information. Thus, we can compare results from these two periods in this figure.

These experiments confirm the utility of our nonlinear aggregation mechanism for making good forecasts of uncertain outcomes. This nonlinear function applies the predictions of a group of people whose individual risk attitudes can be extracted by making them participate in an information market. Equally important, our results show that many of the shortcomings associated with information markets can be bypassed by this two stage method, without having to design and resort to complicated market games. In this context it is worth pointing out that even with such small groups we were able to obtain information whose accuracy, measured by Kullback-Leibler, surpasses by a factor of seven even more complicated institutions such as pari-mutuel games [17].

Lastly, unlike the standard information aggregation implied by the Condorcet theorem, our mechanism allows us to extract probability distributions rather than the validity of a discrete choice obtained via a majority vote. Moreover, our mechanism provides a signal even in situations when an overall system itself does not contain accurate information as to the outcome. Equally important, unlike Condorcet our two-stage mechanism does not demand risk neutrality and access to the same information by all participants in the system.

**Conclusions**

Accurate predictions are essential to individuals and organizations. For large communities, information relevant to forecasts is often dispersed across people, frequently in different geographical areas. Examples include forecasting sales of a product, aggregating the financial predictions of the venture capital community, and public opinion polls. The methodology described in this paper addresses many of the needs to aggregate this information accurately and with the correct incentives. One can take past predictive performance of participants in information markets and create weighting schemes that will help predict future events, even if they are not the same event on which the performance was measured. Furthermore, our two-stage approach can improve upon predictions by harnessing distributed knowledge in a manner that alleviates problems with low levels of participation. The typical business forecast cycle also lends itself to this approach. Since forecasts cycles in organizations typically involve the prediction of similar events on a periodic basis, it is possible to set up an initial market to obtain consistent measures of skills and risk attitudes and then use the reporting mechanism to extract and aggregate information in the future.

Obviously, this approach can also be extended to work across organizations. One possible use is to aggregate and create consensus estimates in the financial analyst community. Another one is to provide the venture capital community a way of forming predictions about the viability of new ventures. The Hollywood Stock Exchange has shown that information markets can be used to predict movie ticket sales, which are tremendously important to studio executives. In the same vein, our methodology can be used with smaller groups of movie screen test subjects to create forecasts before a movie is released. One can imagine a world in which focus groups are no longer run solely on survey questions and discussions, but where each member has a financial stake in the information coming out of the focus group.

The rapid advances of information technologies and the understanding of information economics have opened up many new possibilities for applying mechanism design to gather and analyze information. This paper discusses one such design and provides empirical evidence about its validity. Although the results we presented are particular to events with finite number of outcomes and assumptions of independent information, they can be generalized to continuous state space and non-independent information structure. We are currently pursuing these extensions. Equally intriguing is the possibility of having this mechanism in the context of the Web, thus enabling information aggregation over large geographical areas, perhaps asynchronously. This leads to issues of information cascades and the optimal time to keep an aggregation market open, which we will explore in turn.

**Appendices**

**Appendix 1: Conditional Probabilities and Products of Reports**

**Lemma:** If:
- $O_1$ through $O_n$ are independent observations conditioned on a given state
- The a priori beliefs of the probabilities of the states are uniform

Then $P(s|O_1, O_2, ... O_n) = \dfrac{\prod_{i=1}^{n} P(s|O_i)}{\sum_{s'} \prod_{i=1}^{n} P(s'|O_i)}$ .

In other words, if *N* people observe independent information about the likelihood of a given state and they report those probabilities, one can find the probability conditioned on all of their observations by multiplying their reported probabilities and then normalizing the results.

**Proof (by induction):**

For *N*=1, $P(s|O) = \dfrac{P(s|O)}{\sum_{s'} P(s'|O)} = \dfrac{P(s|O)}{1} = P(s|O)$.

Assume it is true for *N*.

By independence, $P(s|O_1, O_2, ... O_n) = \dfrac{P(s|O_1, O_2, ..., O_n)}{\sum_{s'} P(s'|O_1, O_2, ..., O_n)}$

By Lemma 2 below, $P(s|O_1, O_2, ... O_{n+1})$

$$= \frac{P(s|O_1, O_2, \ldots, O_n) P(s|O_{n+1})}{\sum_{s'} P(s'|O_1, O_2, \ldots, O_n) P(s'|O_{n+1})}$$

$$= \frac{\dfrac{\prod_{i=1}^{n} P(s|O_i)}{\sum_{s'} \prod_{i=1}^{n} P(s'|O_i)} P(s|O_{n+1})}{\sum_{s'} \dfrac{\prod_{i=1}^{n} P(s'|O_i)}{\sum_{s''} \prod_{i=1}^{n} P(s''|O_i)} P(s'|O_{n+1})}$$

$$= \frac{\prod_{i=1}^{n} P(s|O_i) P(s|O_{n+1})}{\sum_{s'} \prod_{i=1}^{n} P(s'|O_i) P(s'|O_{n+1})}$$

$$= \frac{\prod_{i=1}^{n+1} P(s|O_i)}{\sum_{s'} \prod_{i=1}^{n+1} P(s|O_i)} \qquad \blacksquare$$

**Appendix 2: Risk Neutrality and Log Payoff Functions in the Reporting Game**

Consider the following game:
- There are *N* possible states of the world.
- A player is given information about the state of the world $x \in \{1,2,\ldots,N\}$. His belief on the probabilities of these states of the world, conditioned on his information, are $P_i$, $i \in \{1,2,\ldots,N\}$
- The player is asked to report a vector $\{q_1, q_2, \ldots q_N\}$ with the constraint $\sum_{i=1}^{N} q_i = 1$. Then the true state *x* is revealed and he is paid $f(q_x)$.

**Lemma 2:** If the player is risk neutral and $f(y) = log(y)$, then $q_i = P_i$ for all *i*. That is, players will report their true beliefs on the probabilities.

**Proof:**

The player's maximization problem is: $\underset{\{q_i\}}{Max} \sum_{i=1}^{N} P_i \log(q_i)$ s.t. $\sum_{i=1}^{N} q_i = 1$.

The Langrangian for this problem is $L = \sum_{i=1}^{N} P_i \log(q_i) - \lambda \left( \sum_{i=1}^{N} q_i - 1 \right)$

The first order condition is: $\dfrac{P_i}{q_i} = \lambda$ for all $i$  => $P_i = \lambda q_i$

Summing over all $i$, $1 = \lambda$. Thus $q_i = P_i$ for all $i$. ∎